\documentclass[11pt,a4paper]{article}
\pdfoutput=1
\usepackage{jheppub}

\usepackage{amsmath}
\usepackage{verbatim}
\usepackage{amssymb}
\usepackage{inputenc}
\usepackage{textcomp}
\usepackage{appendix}

\newcommand{\e}{\epsilon}
\newcommand{\be}[1]{\begin{equation}\label{#1} }
\newcommand{\ee}{\end{equation}}
\newcommand{\bea}[1]{\begin{eqnarray}\label{#1} }
\newcommand{\eea}{\end{eqnarray}}
\newcommand{\p}{\partial}
\newcommand{\refb}[1]{(\ref{#1})}
\newcommand{\N}{{\mathcal{N}}}

\renewcommand{\L}{{\mathcal{L}}}

\newcommand{\Q}{\mathcal{Q}}
\newcommand{\J}{\mathcal{J}}
\newcommand{\bL}{\bar{{\mathcal{L}}}}
\newcommand{\bQ}{\bar{\mathcal{Q}}}

\renewcommand{\>}{\rangle}
\newcommand{\<}{\langle}

\renewcommand{\a}{\alpha}
\newcommand{\ta}{\tilde{\alpha}}

\newcommand{\C}{\tilde{C}}
\renewcommand{\b}{\beta}
\renewcommand{\t}{\tau}
\newcommand{\s}{\sigma}

\title{Tensionless Superstrings: View from the Worldsheet}

\author[a, b]{Arjun Bagchi} \author[c]{Shankhadeep Chakrabortty} \author[b,d]{and Pulastya Parekh}
\author{\\}
\affiliation[a]{Center for Theoretical Physics, Massachusetts Institute of Technology,\\ 77 Massachusetts Avenue, Cambridge, MA 02139, USA.\\} 

\affiliation[b]{Indian Institute of Technology Kanpur, Kanpur 208016. INDIA.\\} 

\affiliation[c]{Van Swinderen Institute for Particle Physics and Gravity, University of Groningen, \\ Nijenborgh 4, 9747 AG Groningen, The Netherlands.\\ }

\affiliation[d]{Indian Institute of Science Education and Research\\ Dr Homi Bhabha Road, Pashan. Pune 411008. INDIA.\\} 

\emailAdd{abagchi@mit.edu, s.chakrabortty@rug.nl, pulastya.parekh@students.iiserpune.ac.in}

\abstract{In this brief note, we show that the residual symmetries that arise in the analysis of the tensionless superstrings in the equivalent of the conformal gauge is (a trivial extension of) the recently discovered 3d Super Bondi-Metzner-Sachs algebra, discussed in the context of asymptotic symmetries of 3d Supergravity in flat-spacetimes. This helps us uncover a limiting approach to the construction of the tensionless superstring from the point of view of the worldsheet, analogous to the one we had adopted earlier for the closed tensionless bosonic string.}

\preprint{MIT-CTP-4816}

\begin{document}
\maketitle

\section{Introduction}
String theory is the most viable of the current theories of quantum gravity. The fundamental objects of string theory are vibrating strings and the only free parameter in the non-interacting theory is set by the tension of the fundamental string which is given by 
\be{}
T = \frac{1}{4 \pi \alpha'}.
\ee 
In the limit where the tension becomes infinite or $\alpha' \to 0$, the fundamental string shrinks to a point-particle and superstring theory is approximated by its poorer cousin, supergravity. In this limit, we recover everything we could have constructed out of usual quantum field theory. This is the point-particle limit of string theory. 

There exists a diametrically opposite limit of string theory, viz. the ultra-stringy limit, which always been a source of great intrigue. This is where the tension, instead of going to infinity as in the point-particle limit, becomes zero, or $\alpha' \to \infty$ and the fundamental string becomes long and floppy. This is the limit that is supposed to capture the very high energy behaviour of string theory and as a result offers a gateway to the very quantum nature of gravity. 

The tensionless limit of string theory has been investigated since the late 1970's starting with the work of Schild \cite{Schild}. Along a different angle, Gross and Mende \cite{Gross:1987kza, Gross:1987ar} investigated the extreme high energy limit of string theory and discovered that the string scattering amplitudes behave in a very simple way in the $\a' \to \infty$ limit and that there were infinitely many linear relations between these scattering amplitudes hinting at the existence of a higher symmetry structure in this limit \cite{Gross:1988ue}. More recently, the tensionless limit of string theory has been linked to the higher spin structures \cite{Sundborg:2000wp, Witten-talk}  and hence to holographic dualities  \cite{Klebanov:2002ja, Sezgin:2002rt, Gaberdiel:2010pz} where higher spin Vasiliev theories \cite{Vasiliev:2004qz} play a central role. 

\subsection*{Our Point of View: Worldsheet Symmetries}

Our perspective in this field has been somewhat different from the recent focus on higher spin theories, although the initial goal was to concretize this connection between tensionless limits and higher spins (for work in this direction see e.g. \cite{{Chang:2012kt, Gaberdiel:2014cha}}). We have been interested in investigating the tensionless symmetries from the point of view of the world sheet following earlier work \cite{Isberg:1993av}. For a non-exhaustive list of other relevant  work on tensionless (super)strings the reader is referred to \cite{Sundborg:1991ztl}--\cite{Bakas:2004jq}.

One of the central reasons why a lot of progress has been possible in string theory is the emergence of 2 dimensional conformal symmetry on the world-sheet. In the conformal gauge, there is a residual gauge symmetry on the world-sheet of the closed bosonic string which is 2d conformal invariance and this enables us to use techniques of 2d conformal field theory to great effect in the quantisation of string theory. 

Following \cite{Isberg:1993av}, one can see that there is a similar story that emerges in the case of the tensionless closed bosonic string in the analogue of the conformal gauge. The residual symmetry algebra that emerges in this case is the 2d Galilean Conformal Algebra (GCA) \cite{Bagchi:2013bga}. The GCA has shown up in different contexts in recent years, viz. in non-relativistic AdS/CFT \cite{Bagchi:2009my}, in the non-relativistic versions of Electrodynamics \cite{Bagchi:2014ysa} and Yang-Mills theories \cite{Bagchi:2015qcw}, and surprising as the asymptotic symmetries of 3d flat space \cite{Barnich:2006av} and hence has been used to construct the notion of holography in Minkowski spacetimes \cite{Bagchi:2010eg,Bagchi:2012xr, Barnich:2012xq}. Taking inspiration from the methods of 2d GCA constructed e.g. in \cite{Bagchi:2009pe}, we would like to address the tensionless limit of string theory. Some initial steps in this direction were taken in \cite{Bagchi:2015nca}. There were some expected results and some interesting surprises as we briefly now remind the reader. 

\subsection*{Quick Recap: The Bosonic Story}

In \cite{Bagchi:2015nca}, we found that there was an intrinsic method to the theory of tensionless strings and there was a limiting story. The intrinsic method was the one where one started with the action of the tensionless string found in \cite{Isberg:1993av} and assumed that was the defining feature of the tensionless string which was now a fundamental object and not a derived concept. One could then write down the equations of motion, constraints, mode expansions and proceed to quantize the system. In the classical regime, everything, including mode expansions, constraint algebra etc. could be derived by looking at the appropriate limit on the world sheet. 

The above described limit is, interestingly, an ultra-relativistic limit on the world-sheet co-ordinates $(\s, \t)$, viz. $\{ \s \to \s, \t \to \e \t$ with $\e \to 0 \}$. This sends the world-sheet speed of light to zero. For details of this procedure, we encourage the interested reader to have a look at \cite{Bagchi:2013bga, Bagchi:2015nca}. The apparent dichotomy between an ultra-relativistic limit and naming the residual symmetry the Galilean conformal symmetry is resolved when one understands that in the special case of two dimensions, the non-relativistic contraction ($\{ \s \to \e \s, \t \to \t$ with $\e \to 0 \}$) and the above mentioned ultra-relativistic limit yields the same algebra from the two copies of the Virasoro algebra of the tensile string. 

The fact that the limit on the world-sheet is an ultra-relativistic one instead of a non-relativistic one, has intriguing consequences on the physics in the quantum regime. The ultra-relativistic limit mixes the creation and annihilation operators of the tensile string and hence the vacua for the tensile theory and the fundamentally tensionless theory are very different. Bogolioubov transformations on the worldsheet link the two sets of oscillators and the tensionless vacuum becomes expressible as a squeezed state in terms of the tensile oscillators and the tensile vacuum. The tensionless vacuum is thus a very highly energised state in terms of the variables of the tensile theory. 

In \cite{Bagchi:2015nca}, we remarked that our analysis was particularly suited to the study of Hagedorn physics. Strings near the Hagedorn temperature become effectively tensionless and we equated the emergence of the long string in the Hagedorn phase to the emergence of the fundamentally tensionless vacuum on the worldsheet. Contrary to the usual folklore about the destruction of the string worldsheet picture in the vicinity of the Hagedorn temperature, what our analysis predicted is that the Riemannian structure of the tensile world-sheet is replaced by the Carrollian structure of the tensionless worldsheet \cite{Duval:2014lpa}, where a worldsheet metric is no longer the correct variable to describe the worldsheet manifold. This Carrollian structure is very similar (and actually dual) to the Newton-Cartan structures arising in non-relativistic systems. We made preliminary remarks about the thermal nature of the emergent tensionless vacuum by invoking left-right entanglement on the string worldsheet. 

\subsection*{Present considerations}

The obvious generalisation of this formulation is to the tensionless superstring and this is the aim of the current paper. In this short report, we will show that the residual symmetry that emerges on the worldsheet of the tensionless superstring is a particular supersymmetric extension of the GCA. There can be several ways to supersymmetrize the bosonic GCA. The contraction of the two copies of the Virasoro algebra generates a unique answer when we look at non-trivial contractions{\footnote{This statement is somewhat ambiguous. What we mean is that we are not trivially scaling away parts of the algebra. A concrete requirement is that there should be at least one Virasoro left over after the scaling.}}. But there are non-unique ways to contract the supersymmetry generators. It is a priori not clear which one of these should be important to the tensionless superstring. Hence to begin with, in Sec. 2, we visit the algebraic aspects of the Super GCA and mention two different and acceptable contractions of the Super Virasoro which lead to different algebras. We spend some time on the representation theory aspects of the two algebras. 

We then focus our attention on the tensionless superstring in Sec.~3. Here we first look at the tensionless action following \cite{Sundborg:1991ztl} and fix a convenient gauge. We then do the analysis of residual gauge symmetries and to our surprise, we find that the algebra arising from the analysis is a trivial extension of the one recently obtained in \cite{Barnich:2014cwa} which was obtained by looking at the canonical analysis of asymptotic symmetries of 3D supergravity on flat spacetimes. This is one of the two contractions that we considered in Sec.~2. We also perform a mode expansion of the tensionless superstring to find the form of the same symmetry structure. We comment on central extensions and the other contraction performed in Sec.~2. In Sec.~4, we show how one can obtain some of the aspects of the tensionless superstring from the corresponding quantities of the tensile superstring. In Sec.~5, we take some preliminary steps in the construction of the quantum tensionless superstring. We conclude with a summary of our results and a discussion of open directions in Sec.~6. 

\subsection*{Note Added}
As this paper was being readied for submission, the pre-print \cite{Casali:2016atr} appeared on the arXiv containing some overlap with the current paper.

\section{Supersymmetrizing 2d GCA}

We begin by looking at the algebraic aspects of the symmetry that would play a central role in the construction of the tensionless superstring. As we mentioned in the introduction, in the usual bosonic closed string theory, conformal symmetry arises on the worldsheet as a residual symmetry after fixing the conformal gauge in the tensile string action. The 2d Galilean Conformal Algebra similarly arises in the theory of tensionless closed bosonic strings. The tensionless limit on the string worldsheet, as mentioned before, can be understood as an ultra-relativistic contraction. So now that we are interested in the tensionless superstring, we will focus our attention on ultra-relativistic contractions of the tensile superstring algebra, the Super-Virasoro algebra. Before going ahead, we would like to mention previous work on supersymmetrizing the GCA in various dimensions. For general dimensions, this was done in \cite{deAzcarraga:2009ch, Sakaguchi:2009de, Bagchi:2009ke}. For $d=2$, details were worked out in \cite{Mandal:2010gx} following closely the bosonic analysis of \cite{Bagchi:2009pe}. For a recent study of an extended version of SGCA, the reader is referred to \cite{Mandal:2016lsa} 

\subsection*{Two different algebras from contraction}
The residual symmetry algebra of the superstring after fixing the conformal gauge is simply the 2D $\N=(1,1)$ superconformal algebra
\bea{} 
&& [\L_n, \L_m] = (n-m) \L_{n+m} + \frac{c}{12} \, (n^3 -n) \delta_{n+m,0} \cr
&& [\L_n, \Q_r] = \left(\frac{n}{2} -r\right) \Q_{n+r}, \quad \{ \Q_r, \Q_s \} = 2 \L_{r+s} + \frac{c}{3} \, \left(r^2 - \frac{1}{4}\right) \delta_{r+s, 0}.
\eea
and similarly for generators $\bL_n, \bQ_r$ with $c$ replaced with $\bar{c}$. We already know that the bosonic part of the algebra needs to contract in the ultra-relativistic fashion for consistency with the analysis of the bosonic tensionless closed string theory. This is
\be{vir2gca}
L_n = \L_n - \bL_{-n}, \quad M_n = \e \left( \L_n + \bL_{-n} \right)
\ee
From the point of view of the algebra, there are several ways that one could contract the fermionic generators. Below we list two options. 

\bigskip

\paragraph{\em{Homogenous scaling:}} Here we will scale both the supercharges in a similar fashion. 
\be{}
Q^+_r = \sqrt{\e} \ \Q_r, \quad Q^-_r = \sqrt{\e} \ \bQ_{-r}
\ee
This contraction, together with \refb{vir2gca} leads to the following algebra:
\bea{sgcah}
&& [L_n, L_m] = (n-m) L_{n+m} + \frac{c_L}{12} \, (n^3 -n) \delta_{n+m,0} \nonumber\\
&& [L_n, M_m] = (n-m) M_{n+m} + \frac{c_M}{12} \, (n^3 -n) \delta_{n+m,0} \\
&& [L_n, Q^\a_r] = \Big(\frac{n}{2} - r\Big) Q^\a_{n+r}, \quad \{Q^\a_r, Q^\b_s \} = \delta^{\a\b} \left[M_{r+s} + \frac{c_M}{6} \Big(r^2 - \frac{1}{4}\Big)  \delta_{r+s,0} \right] \nonumber
\eea
In the above, $\a, \b = \pm$. The suppressed commutators are all zero. We have also defined
\be{}
c_L = c - \bar{c}, \quad c_M = \e \left( c + \bar{c} \right).
\ee
Since the bosonic part of this algebra is the Galilean Conformal Algebra, we shall call this Super Galilean Conformal Algebra [Homogenous] or SGCA$_H$. Here it is important to note that there are two different labeling of the fermionic generators depending on boundary conditions. The Neveu-Schwarz sector is characterized by $r \in \mathbb{Z} + \frac{1}{2}$ while the Ramond sector is characterized by integer labeling $r \in \mathbb{Z}$. We shall exclusively be dealing with the NS sector through out this paper. 

Interestingly, contraction of the $\N=(1,0)$ conformal algebra would lead to same algebra (stripped of the $(\a,\b)$ indices). This is the very same algebra that arises from the canonical analysis of asymptotic symmetries of 3D $\N=1$ Supergravity recently performed in \cite{Barnich:2014cwa}. The asymptotic symmetry algebra (ASA) of a certain spacetime is the algebra of allowed diffeomorphisms modded out by the trivial diffeos. This often is just the isometries of the vacuum. But there are famous exceptions, the most noteworthy being the Brown-Henneaux analysis of AdS$_3$ \cite{Brown:1986nw} which leads to two copies of the Virasoro algebra. This can be looked at as the first step towards the AdS/CFT correspondence \cite{Maldacena:1997re}. The two copies of the Virasoro algebra dictates the symmetries of the dual 2d field theory, which in this case is a conformal field theory. The ASA thus characterise dual field theories and if the symmetry structure is infinite dimensional, a lot can be said about the field theory just by symmetry arguments. 

In asymptotically flat spacetimes, there are infinite dimensional ASAs when one considers symmetries at the null boundary. These are called Bondi-Metzner-Sachs algebras after the authors who first discovered them in the context of four dimensional Minkowski spacetime \cite{Bondi:1962px}. In three dimensional spacetimes, the BMS$_3$ algebra is actually isomorphic to the 2d GCA given in \cite{Bagchi:2010eg}. The central terms are $c_L=0, \ c_M= 3/G$ for Einstein gravity \cite{Barnich:2006av}. This can be obtained by looking at the ultra-relativistic limit on AdS$_3$ \cite{Bagchi:2012cy,Barnich:2012aw}. The supersymmetric version of the above asymptotic analysis for 3d Minkowski spacetimes was recently addressed in \cite{Barnich:2014cwa}. The authors worked in the Chern-Simons formulation of the simplest version of Supergravity in 3d, viz the minimal $\mathcal{N}=1$ Supergravity and proposed a consistent set of boundary conditions. They derived the ASA through a canonical formulation and named the algebra the super-BMS$_3$. With indices on the supercharges removed, this is precisely the algebra \refb{sgcah}.  The central terms remain the same as mentioned above for the bosonic Einstein gravity case. In \cite{Barnich:2015sca}, 2d field theories with this symmetry algebra were constructed and were shown to be equivalent to a supersymmetric extension of a flat limit of Liouville theory, earlier constructed in \cite{Barnich:2012rz}. 

We will see in the later sections that this is the algebra of residual gauge symmetries on the tensionless superstring worldsheet. This intriguing connection between flat supergravity and the tensionless superstring would possibly prove to be very useful in learning about both sides of this link. We expect some of the physical insights of \cite{Barnich:2014cwa} to be instrumental in our future explorations of tensionless strings.

\bigskip

\paragraph{\em{Inhomogenous scaling:}} Here we scale the supercharges in a way similar to the bosonic generators. 
\be{inh-sc}
G_r = \Q_r - i \bQ_{-r}, \quad H_r = \e \left(\Q_r + i \bQ_{-r} \right)
\ee
This leads to the algebra 
\bea{sgcai}
&& [L_n, L_m] = (n-m) L_{n+m} + \frac{c_L}{12} (n^3 -n) \delta_{n+m,0}, \nonumber\\
&& [L_n, M_m] = (n-m) M_{n+m} + \frac{c_M}{12} (n^3 -n) \delta_{n+m,0}, \\
&& [L_n, G_r] = \Big(\frac{n}{2} -r\Big) G_{n+r}, \ [L_n, H_r] = \Big(\frac{n}{2} -r\Big) H_{n+r}, \ [M_n, G_r] = \Big(\frac{n}{2} -r\Big) H_{n+r}, \nonumber\\
&& \{ G_r, G_s \} = 2 L_{r+s} + \frac{c_L}{3} \Big(r^2 - \frac{1}{4}\Big)   \delta_{r+s,0}, \ \{ G_r, H_s \} = 2 M_{r+s} + \frac{c_M}{3} \Big(r^2 - \frac{1}{4}\Big)   \delta_{r+s,0}.\nonumber
\eea
Again as before, the suppressed commutators are all zero. This algebra is richer than the one we obtained above by the homogeneous contraction of the fermionic generators, the SGCA$_H$ \refb{sgcah}. We shall call this Super Galilean Conformal Algebra [Inhomogenous] or SGCA$_I$. 

An interesting feature of the SGCA$_I$ is that when the central term $c_M$ vanishes, one can show that the algebra truncates to a single copy of the Super Virasoro algebra. The analysis essentially follows the bosonic version outlined in \cite{Bagchi:2009pe}. 

We would like to mention that we are rather intrigued by the appearance of the factor of $i$ in the scaling \refb{inh-sc}. This seems to indicate a fundamental difference between the left-movers and right-movers in the SUSY extension of the Virasoro algebra when we consider the ultra-relativistic limit. This mysterious factor of $i$ is also reminiscent of another such appearance of an imaginary factor in the non-relativistic limit on the bosonic tensionless string oscillators, discussed in Appendix B of our earlier work \cite{Bagchi:2015nca}. We hope to comment on this in future work. 

\bigskip

\paragraph{\em{Non Relativistic Isomorphisms:}} 
One of the very surprising aspects of the contractions of the two copies of the Virasoro algebra is that the ultra-relativistic contraction described above \refb{vir2gca} and the non-relativistic contraction 
\be{}
L_n = \L_n + \bL_{n}, \quad M_n = \e \left( \L_n - \bL_{n} \right)
\ee
both yield the 2d GCA
\bea{sgcai}
&& [L_n, L_m] = (n-m) L_{n+m} + \frac{c_L}{12} (n^3 -n) \delta_{n+m,0} \nonumber\\
&& [L_n, M_m] = (n-m) M_{n+m} + \frac{c_M}{12} (n^3 -n) \delta_{n+m,0}, \\
&& [M_n, M_m] = 0. \nonumber
\eea
This curious fact was first noticed in \cite{Bagchi:2010eg}. This isomorphism between the ultra-relativistic and non-relativistic limits is a curious feature of 2 dimensions. In terms of spacetime, there is one contracted and one non-contracted direction in 2d. The ultra-relativistic limit is a contraction of the time direction while the non-relativistic limit is a contraction of the spatial direction and the fact that the algebra is isomorphic under both contractions is an indication that the procedure does not distinguish between what is space and what is time. In higher $d+1$ dimensions, the ultra-relativistic limit leads to only one contracted direction (time) while the non-relativistic limit contracts $d$ spatial directions. The algebras obtained in the process by contracting relativistic conformal algebras are thus different. When one is looking to generalize this procedure to more involved algebras like $\mathcal{W}$ algebras, there are interesting departures from this isomorphism even in two dimensions. The interested reader is referred to \cite{Campoleoni:2016vsh}. 

In this work, our focus is on super-algebras and we discover that there exists isomorphism similar to that of the case of the Virasoro algebra. \refb{sgcah} is isomorphic to the non-relativistic contraction 
\be{}
L_n = \L_n + \bL_{n}, \quad M_n = \e \left( \L_n - \bL_{n} \right) \quad Q^+_r = \sqrt{\e} \ \Q_r \quad Q^-_r = \sqrt{\e}\ \bQ_r
\ee
The identification of the central charges change in the usual way with the non-relativistic ones being identified as
\be{NRc}
{\mbox{Non-Relativistic limit:}} \quad c_L = c + \bar{c}, \quad c_M = \e \left(c - \bar{c}\right) 
\ee
Similarly, \refb{sgcai} is isomorphic to the non-relativistic contraction \cite{Mandal:2010gx}
\be{}
L_n = \L_n + \bL_{n}, \quad M_n = \e \left( \L_n - \bL_{n} \right), \quad G_r = \Q_r + \bQ_{r}, \quad H_r = \e \left( \Q_r - \bQ_{r} \right).
\ee
The identification of the central terms is again given by \refb{NRc}.
\bigskip

\subsection*{Representation theory}
We will briefly explore the representation theory of these two different algebras, concentrating on highest weight representations. We remind the reader that we are going to be restricting ourselves to the NS sector only. We will first label the states with the Cartan of the algebras $(L_0, M_0)$: 
\be{}
L_0 |h_L, h_M\> = h_L |h_L, h_M\>, \quad M_0 |h_L, h_M\> = h_M |h_L, h_M\>.
\ee
Now we construct the notion of SGCA primary states in close analogy with superconformal primaries. These are states ($|h_L, h_M\>_p$) which are annihilated by the positive modes of the algebras in question.
\bea{}
&&\hspace{-1cm} \mbox{SGCA$_H$:} \quad L_n  |h_L, h_M\>_p = M_n  |h_L, h_M\>_p = Q^\pm_r |h_L, h_M\>_p = 0, \quad \forall \ n, r>0  \\
&&\hspace{-1cm} \mbox{SGCA$_I$:} \quad L_n  |h_L, h_M\>_p = M_n  |h_L, h_M\>_p = G_r |h_L, h_M\>_p = H_r |h_L, h_M\>_p = 0.
\eea
As usual, the Hilbert space is constructed on these primary states by acting on them with negative modded operators. The vacuum of the theory is similarly given by
\be{}
L_n |0\> = M_n |0\> = 0, \quad \forall n\ge 1. 
\ee
along with 
\be{}
\mbox{SGCA$_H$:}  \quad Q^\pm_r |0\> =0, \quad \mbox{SGCA$_I$:} \quad G_r |0\> = H_r |0\> = 0, \quad \forall \ r>\frac{1}{2}.
\ee
The important thing here is that since we are looking at the ultra-relativistic contraction, the vacua of the SGCAs would not be the one we find from the limit of the relativistic theory. This is going to play an important role in the construction of the tensionless quantum superstring.

\subsection*{EM Tensor and its superpartner}

The Energy momentum tensor of the Super-Virasoro is given by:
\be{}
T_{cyl} = \sum_n \L_n e^{-inw}, \quad {\bar{T}}_{cyl} = \sum_n \bL_n e^{-inw}
\ee
The superpartner of the EM tensor on the cylinder: 
\be{}
\J_{cyl} = \sum_r \Q_r e^{-irw}, \quad {\bar{\J}}_{cyl} = \sum_r \bQ_r e^{-irw}
\ee
The SGCA EM tensor is the same as the bosonic one:
\bea{EM}
&&T_1 = T_{cyl} - {\bar{T}}_{cyl} = \sum_n \left( L_n + i n \t M_n \right) e^{-in\s},\\ &&
T_2 = \lim_{\e \to 0} \e \left( T_{cyl} + {\bar{T}}_{cyl} \right) = \sum_n  M_n e^{-in\s}.
\eea
The supercurrents for the SGCA are different for the two different algebras. The SGCA$_H$ version is given by:
\be{sEM}
J_1 = \sqrt{\e} \ \J_{cyl} = \sum_r Q^+_r e^{-i r \s}, \quad J_2 = \sqrt{\e} \ {\bar{\J}}_{cyl} = \sum_r   Q^-_r  e^{-i r \s}
\ee
On the other hand, the SGCA$_I$ version : 
\be{}
J_1 = \J_{cyl} - i{\bar{\J}}_{cyl} = \sum_r (G_r +i r \tau H_r) e^{-i r \s}, \quad J_2 = \e (\J_{cyl} + i{\bar{\J}}_{cyl})= \sum_r   H_r e^{-i r \s}
\ee
The EM tensor and its superpartner will play a crucial role in the analysis of the constraints of the tensionless superstring that we will now go on to describe. 

\newpage

\section{Classical Tensionless Superstrings}

We start this section by reminding the reader of the tensionless closed bosonic string action. This is given by
\be{tlessbos}
S_B = \int d^2 \xi \ V^a V^b \gamma_{ab}
\ee
where $\gamma_{ab} = \p_a X^\mu \p_b X^\nu \eta_{\mu\nu}$ is the induced metric and $V^a$ are weight $1/2$ vector densities which are linked to the metric of the string worldsheet in the tensile theory by the following relation
\be{}V^a V^b = T\sqrt{-g} g^{ab}. \ee
In the case of the tensionless string, the metric degenerates and one needs to replace it by the vector density. The action \refb{tlessbos} is derived by taking the phase space action, integrating out momenta and taking the tension to zero limit \cite{Isberg:1993av}. A similar process (without sending the tension to zero) results in the Polyakov action for the usual tensile string theory. 

\subsection*{Action, equations of motion and residual symmetries}

Our interest in this present work is to look at the tensionless closed superstring. Following the procedure outlined in \cite{Sundborg:1991ztl}, one can supersymmetrise the tensionless closed bosonic string. The action for the tensionless limit of the superstring is given by:
\be{tss}
S =\int d^2\xi \Big[(V^a\p_a X^\mu+i{\chi}\psi^\mu)\cdot(V^b\p_b X^\mu+i{\chi}\psi^\mu)+iV^a{\psi}^\mu\p_a\psi_\mu \Big].
\ee
In the above action, the fermions are densities of weight $-1/4$. The fermionic partner of $X^\mu$ is $\psi^\mu$ and that of $V^a$ is $\chi$, which is connected to the ordinary gravitino $\chi_a$ by the relation $\chi=V^a\chi_a$. 
It is straight-forward to obtain the equations of motion from the above action:
\begin{subequations}\label{eom-tss}
\bea{}
&&\p_a(V^aV^b\p_bX_\mu)+i\p_a(V^aV^b\chi_b\psi_\mu)=0, \\ 
&&i\p_a(V^a{\psi}_\mu)-2V^aV^b(i{\chi}_b\p_aX_\mu+{\chi}_a{\chi}_b\psi_\mu)=0, \\
&&V^b(\p_a X^\mu\p_bX_\mu+2i{\chi}_b\psi^\mu\p_aX_\mu-{\chi}_a\psi^\mu{\chi}_b\psi_\mu)+\frac{i}{2}{\psi}^\mu \p_a\psi_\mu=0, \\
&& V^aV^b\psi^\mu(i\p_bX_\mu-{\chi}_b\psi_\mu)=0.
\eea 
\end{subequations}
As in the usual tensile case, there are gauge symmetries associated with the action \refb{tss}, which is invariant under worldsheet diffeomorphisms and supersymmetry transformations. Transformations of the fields under diffeomorphism, parameterized by $\xi^a$ are
\begin{subequations}\label{}
\bea{}
\delta_\xi X^\mu&=&-\xi^b\p_bX^\mu, \\
\delta_\xi \psi^\mu&=&-\xi^b\p_b\psi^\mu, \\
\delta_\xi V^a&=&-V^b\p_b \xi^a+\xi^b\p_bV^a+\frac{1}{2}(\p_b \xi^b)V^a, \\
\delta_\xi \chi&=&-\xi^a\p_a\chi+\frac{1}{4}(\p_a \xi^a)\chi,
\eea
\end{subequations}
and transformations under %$\mathcal{N}=(1,1)$ 
supersymmetry, parameterized by $\e$ are
\begin{subequations}\label{}
\bea{} 
\delta_\e X^\mu&=& i\e \psi^\mu, \\
\delta_\e \psi^\mu&=& -\e V^a\p_a X^\mu-\frac{1}{2}i\e(\psi^\mu\chi), \\
\delta_\e V^a&=& iV^a(\e\chi_b), \\
\delta_\e\chi&=& V^a\p_a\e.
\eea
\end{subequations}
In light of these transformations, we need to fix a gauge. A convenient choice where physical interpretations become clear is the following gauge:
\be{}
V^a=(1, 0),\ \chi=0.
\ee
The last two equations of motion \refb{eom-tss} become constraints the string is subject to. 
In this gauge, the action becomes:
\be{} 
S=\int d^2\xi \Big[\dot{X}^2+i{\psi}\cdot\dot{\psi}\Big] 
\ee
The equations of motion and the constraints simplify considerably. The equations of motion become
\be{eom}
\ddot{X}^\mu=0, \quad \dot{\psi}^\mu=0.
\ee
The constraints take the form: 
\be{}
\dot{X}^2=0, \quad \dot{X}\cdot X'+\frac{i}{2}\bar{\psi}\cdot\psi'=0, \quad \psi\cdot\dot{X}=0.
\ee
Then we are left with the following differential equations for $\xi$ and $\e$:
\be{}
\dot{\xi^0}=(\xi^1)', \quad \dot{\xi^1}=0, \quad \dot{\e^\a}=0.
\ee
Solving for the parameters give us the form for the allowed diffeomorphism and supersymmetry transformations:
\be{parameter}
\xi^0 = f'(\sigma)\tau+g(\sigma), \quad \xi^1=f(\sigma), \quad \e^\pm = \e^\pm(\sigma).
\ee
This form of the parameters agree with the results of \cite{Gamboa:1989px}. To explicitly see the form of the generators for such a transformation $(\xi,\e)$, we need to consider tranformations in $\mathcal{N}=(1,1)$ superspace. Thus let us consider Grassmanian coordiantes $\theta$ and $\bar{\theta}$ along with the 2 dimensional worldsheet coordinates $\sigma^a$. A general superspace transformation on the worldsheet  is given by
\begin{subequations}\label{} 
\bea{}
\sigma^a&\rightarrow&\sigma'^a=\sigma^a+\xi^a+\frac{i}{2}\Big(\e^+\rho^a\theta+\e^-\rho^a\bar{\theta}\Big), \\
\theta&\rightarrow&\theta'=\theta+\e^++\frac{1}{2}\theta\p_+\xi^+, \\
\bar{\theta}&\rightarrow&\bar{\theta}'=\bar{\theta}+\e^-+\frac{1}{2}\theta\p_-\xi^-. 
\eea
\end{subequations}
However, when we consider the tensionless limit, the worldsheet metric $\eta^{ab}$ becomes degenerate. Thus the interpretation of the Clifford Algebra : $\{\rho^a,\rho^b\}=2\eta^{ab}$ needs to be changed. The metric $\eta^{ab}$ is replaced by a product of the vector densities $V^a V^b$ (subject to a gauge choice of $V^a=\{1,0\})$. Hence the Clifford Algebra is modified to 
\be{} \{\rho^a,\rho^b\}=2V^aV^b. \ee 
A convenient choice of the gamma matrices $\rho^a$ to satisfy this algebra, is to replace them by the vector density $V^a$, as $\{V^a,V^b\}=2V^aV^b$. %They are represented by $1\times 1$ matrices in the process.
This is also manifested by the choice of the action \refb{tss}, where $V^a$ assumes the role of $\rho^a$ matrices in the fermionic part of the action. %, when compared to its tensile counterpart.
%This is also manifested by the choice of the action \refb{tss} for the tensionless superstring, where the $\rho^a$ matrices in the fermionic part in the tensile case, becomes $V^a$ in the tensionless case.
We recommend the reader to see \cite{Sundborg:1991ztl} for further details. The transformation on superspace thus becomes
\begin{subequations}\label{}
\bea{}
\sigma^a&\rightarrow&\sigma'^a=\sigma^a+\xi^a+\frac{i}{2}\e^+ V^a\theta+\frac{i}{2}\e^-V^a\bar{\theta}, \\
\theta&\rightarrow&\theta'=\theta+\e^++\frac{1}{4}\theta\p_a\xi^a, \\
\bar{\theta}&\rightarrow& \bar{\theta}'=\bar{\theta}+\e^-+\frac{1}{4}\bar{\theta}\p_a\xi^a.
\eea
\end{subequations}

A superfield $\Phi(\sigma^a,\theta,\bar{\theta})$ would transform as
\be{sfield}
\delta\Phi=(\delta\tau\p_\tau+\delta\sigma\p_\sigma+\delta\theta\p_\theta+\delta\bar{\theta}\p_{\bar{\theta}})\Phi.
\ee
Once we have chosen $V^a=(1,0)$, we could expand \refb{sfield} with the appropriate forms of $\xi$ and $\e$ from \refb{parameter}
\begin{subequations}
\bea{}
\delta\Phi&=&\Big[\Big(f'\tau+g+\frac{i}{2}\e^+\theta+\frac{i}{2}\e^-\bar{\theta}\Big)\p_\tau+f\p_\sigma \nonumber
\\&&\hspace{2cm}+\Big(\e^++\frac{1}{2}f'\theta\Big)\p_\theta+\Big(\e^-+\frac{1}{2}f'\bar{\theta}\Big)\p_{\bar{\theta}}\Big]\Phi \\
%&=[(f'\tau\del_\tau+f\del_\sigma+\frac{1}{2}f'\theta\del_\theta+\frac{1}{2}f'\bar{\theta}\del_{\bar{\theta}})+g\del_\tau+\e^+(\del_\theta+\frac{i}{2}\theta\del_\tau)+\e^-(\del_{\bar{\theta}}-\frac{i}{2}\bar{\theta}\del_\tau)]\Phi. \\
&=&[L(f)+M(g)+Q^+(\e^+)+Q^-(\e^-)]\Phi ,
\eea
\end{subequations}
where the generators have the form
\begin{subequations}
\bea{} 
L(f)&=&f\p_\sigma+f'\Big[\tau\p_\tau+\frac{1}{2}\Big(\theta\p_\theta+\bar{\theta}\p_{\bar{\theta}}\Big)\Big] ;\quad M(g)=g\p_\tau \\
Q^+(\e^+)&=&\e^+\Big(\p_\theta+\frac{i}{2}\theta\p_\tau\Big);\quad Q^-(\e^-)=\e^-\Big(\p_{\bar{\theta}}+\frac{i}{2}\bar{\theta}\p_\tau\Big). 
\eea
\end{subequations}
Here $L$ and $M$ are the bosonic generators, while $Q^\pm$ are fermionic generators. Since all the parameters $f,g$ and $\e^\pm$ are functions of $\sigma$ we can do a fourier expansion in $e^{in\sigma}$ to each of them as follows: %, and obtain the fourier modes of the generators $L_n,M_n$ and $Q^\pm_n$. \\
\be{}
f=\sum_n a_ne^{in\sigma};\quad g=\sum_n b_ne^{in\sigma};\quad \e^\pm=\sum_n \zeta^\pm_ne^{in\sigma}. \\
\ee
Defining the bosonic and fermionic generators such that
\be{}
L(f)=-i\sum_n a_nL_n;\quad M(g)=-i\sum_n b_nM_n;\quad Q^\pm(\e^\pm)=\sum_n \zeta^\pm_nQ^\pm_n, \\
\ee
we will obtain the fourier modes of the generators 
\bea{}
L_n&=&ie^{in\sigma}\Big[\p_\sigma+in\tau\p_\tau+\frac{in}{2}\Big(\theta\p_\theta+\bar{\theta}\p_{\bar{\theta}}\Big)\Big];\quad M_n=ie^{in\sigma}\p_\tau; \\
Q^+_n&=&e^{in\sigma}\Big(\p_\theta+\frac{i}{2}\theta\p_\tau\Big);\quad Q^-_n=e^{in\sigma}\Big(\p_{\bar{\theta}}+\frac{i}{2}\bar{\theta}\p_\tau\Big)
\eea
satisfying the following algebra
\bea{}
&& [L_m,L_n]=(m-n)L_{m+n};\quad[L_m,M_n]=(m-n)M_{m+n}; \\
&& [L_m,Q^\a_r]=\left(\frac{m}{2}-r\right)Q^\a_{m+r};\quad\{Q^\a_r,Q^{\a'}_s\}=\delta^{\a\a'}M_{r+s}; \\
&& [M_m,M_n]=[M_m,Q^\a_r]=0.
\eea
This is the 2d SGCA$_H$ that we discussed in the previous section, without the central extensions. When we move to the quantum tensionless superstring, the algebra would get central extensions as indicated in the previous section. 

\subsection*{Mode expansions}
We will now look to derive the above symmetries in a different way, viz. by solving the equations of motion \refb{eom-tss} and using mode expansions. The solutions in the NS-NS sector are of the following form
\bea{NSsoln}
X^\mu(\tau,\sigma)&=&x^\mu+\sqrt{2c'}B^\mu_0\tau+i\sqrt{2c'}\sum_{n\neq 0}\frac{1}{n}(A^\mu_n-in\tau B^\mu_n)e^{-in\sigma} \\
\psi_\pm^\mu(\sigma,\tau)&=&\sqrt{2c'}\sum_r \beta^{\mu\pm}_r e^{-ir\sigma}
\eea
The commutation relations follows by considering: 
\be{modecom}
[X^\mu(\sigma),\dot{X}^\nu(\sigma')]=\eta^{\mu\nu}\delta(\sigma-\sigma'), \quad \{ \psi^\mu_\a(\sigma),\psi^\nu_{\a'}(\sigma') \}=\eta^{\mu\nu}\delta_{\a\a'}\delta(\sigma-\sigma') 
\ee
and that of the modes:
\be{abb}
[A_m^\mu,B_n^\nu] = 2m\delta_{m+n}\eta^{\mu\nu}, \quad \{\beta_{r}^{\mu\a},\beta_{s}^{\nu\a'} \} = \delta^{\a \a'}\delta_{r+s}\eta^{\mu\nu}.
\ee
We note here that the commutation relations of the oscillators are not in a simple harmonic oscillator basis. This would be of importance later. 
The constraints can be worked out by taking the appropriate derivatives with respect to $\tau$ and $\sigma$. Writing them down explicitly gives us
\begin{subequations}\label{const}
\bea{}
\dot{X}\cdot X'+\frac{i}{2}\bar{\psi}\cdot\psi'&=&2c'\sum_{n}\Big[\sum_{m} A_{-m}\cdot B_{m+n}-in\tau \sum_{m} B_{-m}\cdot B_{m+n} \nonumber\\
&&\ \ \ \ \  +\frac{1}{2}\sum_{r}(2r+n)(\beta^+_{-r}\cdot\beta^+_{r+n}+\beta^-_{-r}\cdot\beta^-_{r+n})\Big]e^{-in\sigma} \nonumber\\
&=& 4c'\sum_{n} \Big[L_n-in\tau M_n \Big]e^{-in\sigma}=0,\\
\dot{X}^2&=&2c'\sum_{n}\Big[\sum_{m} B_{-m} \cdot B_{m+n}\Big]e^{-in\sigma}=4c'\sum_{n}M_ne^{-in\sigma}=0, \\
\psi_\pm\cdot\dot{X}&=&2c'\sum_{r}\Big[\sum_{m} B_{-m}\cdot \beta^\pm_{m+r}\Big]e^{-ir\sigma}=4c'\sum_{r}Q^\pm_r e^{-ir\sigma}=0.
\eea
\end{subequations}
In the above, we have made the following definitions:
\bea{}
L_n&=&\frac{1}{2}\sum_{m}  A_{-m}\cdot B_{m+n}+\frac{1}{4}\sum_{r}(2r+n)\Big(\beta^+_{-r}\cdot\beta^+_{r+n}+\beta^-_{-r}\cdot\beta^-_{r+n}\Big)  \\
M_n&=&\frac{1}{2}\sum_{m} B_{-m} \cdot B_{m+n} \\
Q^\pm_r&=&\frac{1}{2}\sum_{m} B_{-m} \cdot \beta^\pm_{m+r} 
\eea
The classical algebra of the modes of the constraints can be obtained by using the commutation and anti-commutation relations of the oscillators \refb{abb} : 
\bea{SGCA}
&&[L_m,L_n]=(m-n)L_{m+n}, \quad [L_m,M_n]=(m-n)M_{m+n}, \quad [M_m, M_n]=0, \\
&&[L_m,Q^\a_r]=\Big(\frac{m}{2}-r\Big)Q^\a_{m+r}, \quad [M_m,Q^\a_r]=0, \quad \{Q^{\a}_r,Q^{\b}_s\}=\delta^{\a\b}M_{r+s}.
\eea
We have rederived the SGCA$_H$ from the modes of the solutions of the equations of motion. As stressed previously, this is the symmetry algebra without any central extensions and when we look to quantize the theory, this algebra would get extended. These central extensions would then determine the consistency of the tensionless superstring.

\subsection*{Central terms and the other symmetry algebra}

Like in the bosonic case, the limit from the tensile string suggests that these central extensions would be zero hinting at the fact that tensionless superstrings would be consistent in any number of dimensions. Let us just remind the reader of the argument of the vanishing of the central terms in the limit \cite{Bagchi:2013bga}. The algebra has two distinct central extensions: $c_L=c -\bar{c}$ and $c_M= \e (c +\bar{c})$. If we start from a parent theory free from diffeomorphism anomaly, then $c = \bar{c}$ in the parent theory and hence $c_L =0$. Again, $c$ counts the number of bosonic worldsheet fields in the tensile theory and hence in any healthy theory, this is a finite number which does not scale like $\frac{1}{\e}$. So, $c_M=0$. As before, we may wish to make a distinction between tensionless theories as derived from a tensile theory or a fundamentally tensionless theory. The arguments above hold for a derived theory and for a fundamentally tensionless theory, one needs to rework the usual tensile methods to check what restrictions on dimensions there may emerge in this case. 

As mentioned before, when stripped off the $(\a,\b)$ indices, this above symmetry algebra is the very same algebra that arises from the canonical analysis of asymptotic symmetries of 3D $\N=1$ Supergravity recently performed in \cite{Barnich:2014cwa} and is called the Super BMS$_3$.  

It is also of interest to see that the classical constraints above \refb{const} take the form required from the algebra, viz. EM tensor $= 0$. We can compare the equations of \refb{const} with 
\refb{EM} and \refb{sEM} to see how the algebraic considerations considered before fit into this particular example. 

What we have seen here is that if we use the choice of gauge \refb{}, then there is a residual gauge symmetry which is the SGCA$_H$. We would be able to organize the physics of the tensionless superstring in terms of this symmetry algebra. It is conceivable that there exists another ``good" choice of gauge which lands us on SGCA$_I$ as the residual symmetry algebra. This will again be a preferred choice of gauge. 

We saw in the bosonic case in \cite{Bagchi:2015nca}, that when we considered closed tensionless strings with $c_M=0$, the residual gauge symmetry truncated to a single copy of the Virasoro algebra indicating a deep relation between the tensionless closed string and the open string, previously anticipated e.g. in \cite{Sagnotti:2011qp}. If we are able to come up with a choice of gauge such that in the tensionless superstring the residual symmetry becomes SGCA$_I$, there is the obvious advantage of making the supersymmetric version of the relation between the closed and open strings even here. As we have mentioned earlier in Sec 2, with $c_M=0$, the SGCA$_I$ reduces to a single copy of the Super-Virasoro algebra. We have, however, been unable to find this gauge yet. It is also possible that one may be able to make the connection between closed and open tensionless strings even in the current gauge,- but as if often with gauge choices, the physics of this particular property seems to a bit obscured and harder to interpret in the current form.

\section{Limit from the tensile superstring}
In our analysis so far, we have been essentially confined to thinking of the tensionless superstring as an intrinsic object and we have derived the various properties of the tensionless superstring theory without taking recourse to the contraction on the worldsheet. We now wish to venture into the other direction where we make full use of the mapping between the tensile and tensionless theory i.e. by taking the ultra-relativistic contraction of the tensile string worldsheet. 

\subsection*{Tensile case}

Before that, let us briefly review the tensile case \cite{Green:1987sp}.  Starting from the tensile action:
\be{}
S=-\frac{T}{2}\int d^2\sigma \Big[\p_aX^\mu\p^aX_\mu-i\bar{\psi}^\mu\rho^a\p_a\psi_\mu\Big]
 \ee
The equations of motions are:
\be{} 
\p^a\p_a X^\mu=0; \quad \p_\mp\psi^\mu_\pm=0. 
\ee
The bosonic and fermionc mode expansions (in the NS-NS sector) are listed below:
\begin{subequations}
\bea{}
X^\mu(\tau,\sigma)&=&x^\mu+2\sqrt{2\a'}\a^\mu_0\tau+i\sqrt{2\a'}\sum_{n\neq 0}\frac{1}{n}\Big[\a^\mu_ne^{-in(\tau+\sigma)}+\ta^\mu_ne^{-in(\tau-\sigma)}\Big] \\
\psi_+^\mu(\sigma,\tau)&=&\sqrt{2\a'}\sum_{r \in \mathbb{Z} + \frac{1}{2}} b^{\mu}_r e^{-ir(\tau+\sigma)}, \quad \psi_-^\mu(\sigma,\tau)=\sqrt{2\a'}\sum_{r \in \mathbb{Z} + \frac{1}{2}} \tilde{b}^{\mu}_{r} e^{-ir(\tau-\sigma)}
\eea
\end{subequations}
The relations \refb{modecom} are applicable, and gives us the following brackets:
\be{mode}\begin{split}
[\a_m^\mu,\a_n^\nu]&=[\ta_m^\mu,\ta_n^\nu]=m\delta_{m+n}\eta^{\mu\nu} \\
\{b_{r}^{\mu},b_{s}^{\nu} \}&=\{\tilde{b}_{r}^{\mu},\tilde{b}_{s}^{\nu} \}=\delta^{\a\a'}\delta_{r+s}\eta^{\mu\nu} 
\end{split} \ee
 The components of the energy momentum tensor and the super current are directly related to the constraints:
\bea{}
T_{\pm\pm}&=&(\dot{X}\pm X')^2+\frac{i}{2}\psi_\pm\cdot\p_\pm\psi_+=0 \\
%T_{--}&=&(\dot{X}-X')^2+\frac{i}{2}\psi_-\cdot\p_-\psi_-=0 \\
J_\pm&=&\psi_\pm\cdot\p_\pm X =0
\eea
In terms of modes, the Super-Virasoro constraints are given by
\bea{}
\L_n&=&\frac{1}{2}\sum_m \a_{-m}\cdot\a_{m+n}+\frac{1}{4}\sum_r (2r+n)b_{-r}\cdot b_{r+n} \\
\Q_r&=&\sum_m \a_{-m}\cdot b_{m+r} .
\eea
Using \refb{mode} the Super-Virasoro Algebra (without central terms) is generated:
\be{svira}
[\L_m,\L_n] = (m-n)\L_{m+n}, \quad [\L_m,\Q_r] = \Big(\frac{m}{2}-r\Big)\Q_{m+r},\quad \{\Q_r,\Q_s\} = 2\L_{r+s}
\ee

\bigskip

\subsection*{Tensionless contraction}

The tensionless limit on the worldsheet is obtained by taking the following contraction $\tau\rightarrow\e\tau$; $\sigma\rightarrow\sigma$. The physical insight at play here is that in the tensionless limit, the string will grow long and floppy (as opposed to shrinking into a point particle in the $\a \to 0$ limit) \cite{Bagchi:2013bga} and hence the limit to consider would be one where $\s \to \infty$. But we are on the closed string worldsheet and would like to identify $\s \sim \s + 2 \pi$. So, in the case of the closed string, the contraction to consider is the ultra-relativistic limit just mentioned above. Along with this, one needs to specify how the fermions would scale in the case of the tensionless superstring. Since we have the SGCA$_H$, the fermionic scaling would be homogenous: $\psi\rightarrow\sqrt{\e}\psi$.  

Let us now apply this on the mode expansions of the tensile case to see how to obtain the tensionless ones. We need to keep in mind that we also should scale $\a'$ as 
$\a' \rightarrow \frac{c'}{\e}$. The bosonic modes transform under this scaling as 
\bea{}
X^\mu(\tau,\sigma)&=& x^\mu+2\sqrt{\frac{2c'}{\e}}\a^\mu_0\e\tau+i\sqrt{\frac{2c'}{\e}}\sum_{n\neq 0}\frac{1}{n}\Big[\a^\mu_ne^{-in\sigma}(1-i\e n\tau)+\ta^\mu_ne^{in\sigma}(1-i\e n\tau) \Big] \\
&=& x^\mu+\sqrt{2c'}\sqrt{\e}(\a^\mu_0+\ta^\mu_0)\tau+i\sqrt{2c'}\sum_{n\neq 0}\frac{1}{n}\Big[\frac{1}{\sqrt{\e}}(\a^\mu_n-\ta^\mu_{-n})-i n\tau\sqrt{\e}(\a^\mu_n+\ta^\mu_{-n})\Big]e^{-in\sigma} 
\nonumber
\eea
The fermionic ones scale as
\bea{}
\psi_+^\mu(\sigma,\tau)&=&\sqrt{\e}\sqrt{2\a'}\sum_r b^{\mu}_r e^{-ir\sigma}(1-i\e r\tau) \approx\sqrt{2c'}\sum_r b^{\mu}_r e^{-ir\sigma} \\%(1-i\e n\tau) \\
\psi_-^\mu(\sigma,\tau)&=&\sqrt{\e}\sqrt{2\a'}\sum_r \tilde{b}^{\mu}_{r} e^{+ir\sigma}(1-i\e r\tau) \approx \sqrt{2c'}\sum_r \tilde{b}^{\mu}_{-r} e^{-ir\sigma}
\eea
If we compare these modes with the one that we obtained intrinsically, it can be easily seen that
\begin{subequations}\label{oscmap}
\bea{}
&& A^\mu_n =\frac{1}{\sqrt{\e}}\Big(\a^\mu_n-\ta^\mu_{-n}\Big), \quad  B^\mu_n=\sqrt{\e}(\a^\mu_n+\ta^\mu_{-n}), \\
&& \beta^{\mu+}_r=b^{\mu}_r, \quad \beta^{\mu-}_r={\tilde{b}}^{\mu}_{-r}. \label{fermosc}
\eea
\end{subequations}
 Plugging the above relations back into the constraints give us the connection between the tensile and the tensionless constraints. 
\bea{}
L_n&=&\L_n-\bL_{-n}, \quad M_n = \e (\L_n+\bL_{-n}) \\
Q^+_r&=&\sqrt{\e}\Q_r \quad Q^-_r=\sqrt{\e}\bQ_{-r} 
\eea
This scaling, together with \refb{svira}, corresponds to the algebra obtained in \refb{SGCA}. Therefore this analysis agrees with the ``homogeneous scaling'' of the Super-Virasoro algebra to arrive at SGCA$_H$. It is interesting to note here that at the level of the oscillators, the fermionic ones \refb{fermosc} don't get scaled at all in this particular contraction of the parent Super Virasoro algebra to the homogeneous SGCA.

\section{Quantum Tensionless Superstrings}
We have so far dealt exclusively with classical aspects of the tensionless superstring. We now want to make some preliminary remarks about its quantum nature.  

As in the bosonic case, the theory of quantum superstrings is best formulated in the covariant approach by quantising the theory as a free theory and then imposing the constraints as physical 
conditions on the states of the Hilbert space. For the tensionless superstring, in close analogy with the treatment of the bosonic tensionless string, there are two distinct methods in which this process can be approached. In both cases, the tensionless constraints are imposed on the states of the corresponding Hilbert space. These constraints are
\be{}
\< phy'| T_1 |phy\> = 0, \quad \< phy'| T_2 |phy\> = 0,  \quad \< phy'| J_1 |phy\> = 0,  \quad \< phy'| J_2 |phy\> = 0,
\ee
where the classical conditions \refb{const} have been transformed to their quantum counterparts by elevating them to quantum operators and sandwiching them in between physical states $|phy\>, \ |phy' \>$. In terms of the modes of the EM tensor, this boils down to 
\be{phy}
\< phy'| L_n |phy\> = 0, \quad \< phy'| M_n |phy\> = 0, \quad \< phy'| Q^\pm_r |phy\> = 0.
\ee

The crucial point of difference is the fact that the Hilbert spaces of the limiting and the fundamental theories are different. When we wish to discuss the tensionless superstring theory as a sub-sector of a well-behaved tensile superstring theory, the Hilbert space that we would start with is that of the tensile theory and we will impose the above constraints on the states of that theory to obtain the allowed states of the tensionless subsector. This Hilbert space is built out of the tensile vacuum and all the creation operators acting on this vacuum. The vacuum is defined by
\be{}
|0\> : \quad \a_n | 0\> = \ta_n |0\> = 0; \quad b_r | 0\> = \tilde{b}_r | 0\> = 0. 
\ee
So a general state $|\phi \>$ in this Hilbert space is built up of oscillators acting on the vacuum: 
\be{}
|\phi \> =  \prod \a_{-n_1} \ldots \a_{-n_N} \ . \ \ta_{-m_1} \ldots \ta_{-m_M} \ . \ b_{-r_1} \ldots b_{-r_R} \ . \ \tilde{b}_{-s_1} \ldots \tilde{b}_{-s_S} \ | 0\>
\ee
The tensionless constraints acting on these states give us the physical states. The mass is derived from the $M_0$ constraint:
\be{}
M_0 | \phi \> =0 \ \ \Rightarrow \sum_n B_{-n} \cdot B_{n} |\phi\> =0
\ee
Now we have, 
\be{}
P_\mu = \frac{1}{2 \pi c'} \dot{X}_\mu
\ee
which leads to the total momentum $p_\mu$ of the string
\be{}
p_\mu = \int_0^{2\pi} d \s P_\mu = \sqrt{\frac{2}{c'}} B_{0 \ \mu} 
\ee
The mass of the state is given by 
\bea{}
m^2 |\phi\> &=& - p_\mu p^\mu |\phi\> = - \frac{2}{c'} B_0 \cdot B_0 |\phi\> = \frac{2}{c'} \sum_{n \neq 0} B_{-n} \cdot B_{n} |\phi\> \nonumber\\
&=&\lim_{\e \to 0} \frac{2}{c'} \left[ \sum_{n \neq 0} \e ( \a_{-n} + \ta_{n}) (\a_{n} + \ta_{-n})  \prod \a_{-n_1}..\a_{-n_N}.\ta_{-m_1}..\ta_{-m_M}.b_{-r_1}..b_{-r_R} .\tilde{b}_{-s_1}..\tilde{b}_{-s_S} \ | 0\> \right] \nonumber\\
\Rightarrow m^2 |\phi\> &=& 0. \nonumber
\eea
So we get that in the case of the tensionless superstrings, just as in the bosonic case, the tensionless limit from the tensile theory lands up in a subsector which is characterised by states which have zero mass. These states, as is obvious from the spacetime index structure (which we have suppressed above), also can carry arbitrary spins and hence this is a link between the tensionless superstrings and the theory of massless higher spins. 
It is interesting to note that since $M_n$ does not get any contribution from the fermionic modes, the mass formula does not change from the case of the bosonic tensionless sector.

The fundamental tensionless theory is built out of oscillators which are intrinsically tensionless. The oscillators $A, B$ do not obey a harmonic oscillator algebra \refb{abb} and hence we need to change to a different basis \cite{Bagchi:2015nca}. We will call these operators $C, \C$. These are defined as
\be{}
C_n = A_n + B_n, \quad \C_n = - A_{-n} + B_{-n}
\ee 
These new oscillators now do obey the harmonic oscillator algebra. In terms of the tensile oscillators, these are given by 
\be{}
C_n = f_+ \a_n + f_- \ta_{-n}, \quad \C_n = f_- \a_{-n} + f_+ \ta_n, \quad \mbox{where} \ \ f_\pm = \frac{\sqrt{\e}}{2} \pm \frac{2}{\sqrt{\e}}.
\ee
The main point, as stressed above, is that the fundamental tensionless Hilbert space is very different from the tensile theory. The tensionless Hilbert space is built from the tensionless vacuum $|00\>$ which is defined by 
\be{}
|00\>: \quad  C_n |00\> = \C_n |00\> = \b^\pm_r |00\> = 0, \quad n, r>0.
\ee
A general state in the fundamentally tensionless Hilbert space is given by 
\be{}
|\psi \> = \prod C_{-n_1} \ldots C_{-n_N} \ . \ \C_{-m_1} \ldots \C_{-m_M} \ . \ \b^+_{-r_1} \ldots \b^+_{-r_R} \ . \ \b^-_{-s_1} \ldots \b^-_{-s_S} \ |00\>. 
\ee
The allowed states are the ones in this Hilbert space subject to the physical state conditions \refb{phy}. Since the C's are a mixture of tensile creation and annihilation operators and the map between the fermionic oscillators are given by 
\be{}
\b^+_r = b_r, \quad \b^-_r = \tilde{b}_{-r}
\ee
it is clear that the tensionless limit of the tensile vacuum does not land us onto the tensionless vacuum. The two sets of oscillators are linked by Bogolioubov transformations on the worldsheet and the tensionless vacuum can be described as squeezed state in terms of the tensile vacuum and the tensile oscillators, just as described in \cite{Bagchi:2015nca}. Again, since the $M_0$ constraint does not change from the bosonic theory, the mass operator is not diagonal on the allowed states in the tensionless theory. Hence mass does not have a good interpretation on the states of this fundamentally tensionless theory \cite{Bagchi:2015nca}. 

It is important to note here again that we have restricted our attention purely to the NS-NS sector of the tensionless superstring theory. The R sector is more intricate and will be dealt with at length in upcoming work. 

\section{Conclusions}
In this paper, we have looked at the theory of tensionless superstrings following earlier work \cite{Sundborg:1991ztl}. We have seen that when we fix the equivalent of the conformal gauge in the tensionless superstring action, there is a residual gauge symmetry, the algebra of which is given by the homogenous Super Galilean Conformal Algebra or SGCA$_H$. This (stripped of the indices on the supercharges) is actually the same as the algebra obtained recently in the analysis of asymptotic analysis of 3D Supergravity in asymptotically flat spacetimes \cite{Barnich:2014cwa}.

We performed an intrinsic analysis of the tensionless superstrings residual symmetries and rederived the same by looking at the solutions of the equations of motion using mode expansions. We obtained the same from a contraction on the string worldsheet. When we were looking at classical aspects of the theory, the intrinsic and the limit procedure gave us the same answers. We briefly commented on the quantum aspects of the two approaches and showed that the vacua of the two were markedly different and that the usual tensionless limit on the tensile vacuum does not land one on the fundamentally tensionless vacuum. 

In the process, we also visited some algebraic aspects of GCAs and showed two different ultra-relativistic contractions leading to two different supersymmetrizations of the parent bosonic GCA. We commented on some aspects of the representation theory of the two algebras. 

It is important to emphasise that this is just a stepping stone for what is a much longer project of trying to understand tensionless strings from symmetries of the worldsheet. Like in the bosonic case, we have now found an organising principle which will help us systematically solve the theory of tensionless superstrings by appealing to the symmetries on the string worldsheet. There are numerous things to be addressed and some of it is already work in progress. Let us comment on some open directions before we close. 

Most immediately, we would like to expand on our comments on the different vacua of the two different approaches to the tensionless theory, - viz. the fundamental and the limiting one. We have pointed out that the NS-NS vacua are different. We would like to extend our analysis to include the R sector and analyse the structure of the Bogoliubov transformations between the two sets of oscillators. We had also remarked about the thermal nature of the tensionless vacuum in \cite{Bagchi:2015nca}. It is very likely that a similar story would hold true in the supersymmetric case. This will be explored in upcoming work. 

One of the most intriguing connections outlined in \cite{Bagchi:2015nca} was the connection to Hagedorn physics. It was remarkable that the notion of a worldsheet description still existed near the Hagedorn temperature defying earlier notions. In the supersymmetric case, we wish to investigate this in more detail. The bosonic case suggested the use of Carrollian manifolds for this application. We expect Super-Carrollian structures to emerge in the supersymmetric version. 

One of the principal goals of this project is to understand scattering amplitudes in the tensionless limit from the symmetries on the worldsheet. We would obviously need to first understand the bosonic calculation, but the final goal would be to look at the supersymmetric tensionless string theory and the scattering amplitudes there.

Another of our current avenues of exploration is the fate of spacetime in this tensionless limit. String theory (in its second quantised avatar) is supposed to be background independent and is expected to determine the geometry in which it propagates. Hence a singular limit on the string worldsheet would mean that something singular is also happening to the spacetime where it propagates. We are in the process of investigating the consequences of this for the bosonic theory and the obvious generalisation would be to consider the case of the superstring. 

There are five consistent superstring theories and all of them should admit tensionless limits. It would be of interest to examine all of these carefully and also understand the fate of the web of dualities that link the parent tensile theories in the tensionless limit. The departures of the classical and quantum superstrings needs to be examined in each case to see if there is further substructure emerging in these different theories. 

The recently discovered link between the ambitwistors and null strings \cite{Casali:2016atr} is also very intriguing and opens completely new avenues of applications of our methods.  

\bigskip

\section*{Acknowledgements}
It is a pleasure to thank Aritra Banerjee, Rudranil Basu and Shailesh Lal for discussions. We acknowledge the support of various funding agencies: AB thanks the Fulbright Foundation, PP thanks the Council of Scientific and Industrial Research. We also gratefully acknowledge the warm hospitality of many institutes/universities where parts of this work were carried out: the Simon Center for Geometry and Physics, Stony Brook (AB and PP), Center for Theoretical Physics, Massachusetts Institute of Technology (PP), Saha Institute of Nuclear Physics, Kolkata (PP),  Universite Libre de Brussels (AB and SC), LPTHE Paris (AB) and the Albert Einstein Institute, Potsdam (AB). 

\bigskip

\bigskip

%\newpage

\end{document}